\documentclass[preprint,floatfix,preprintnumbers,amsmath,amssymb]{revtex4}

\usepackage{amsmath}
\usepackage{graphicx}

\newcommand{\ud}{\mathrm{d}}

\newcommand{\be}{\begin{equation}}
\newcommand{\ee}{\end{equation}}

\begin{document}

\title{Non-Linear Beam Splitter in Bose-Einstein Condensate Interferometers}

\author{L. Pezz\'e$^{1,2}$, A. Smerzi$^{1,2}$, G.P. Berman$^1$, A.R. Bishop$^1$ and L.A. Collins$^1$}
\affiliation{1) Theoretical Division, Los Alamos National Laboratory,
Los Alamos, New Mexico 87545, USA\\
2) BEC-INFM and Dipartimento di Fisica, Università di Trento, I-38050 Povo, Italy}

\date{\today}

\begin{abstract} 

A beam splitter is an important component of an atomic/optical Mach-Zehnder interferometer.
Here we study a Bose Einstein Condensate beam splitter, 
realized with a double well potential of tunable height.
We analyze how the sensitivity of a Mach Zehnder interferometer is degraded by the non-linear 
particle-particle interaction during the splitting dynamics. 
We distinguish three regimes, Rabi, Josephson and Fock, and 
associate to them a different scaling of the phase sensitivity with the total number of particles. 
\end{abstract}

\maketitle

\section{introduction}

Sub shot-noise interferometric measurements have become
the subject of lively experimental and theoretical studies in view of possible breakthrough technological 
applications (for a recent review see \cite{Giovannetti_2004}).
In particular, interferometry with dilute Bose-Einstein condensates has become an important tool for 
experiments in fundamental and applied physics \cite{Kasevich_2002}. 
Among these, we mention the recent realization of double-slit \cite{Shin_2004, Schumm_2005}
and Michelson-Morley interferometers \cite{Wang_2005}, and the study of Hambury Brown-Twiss effect \cite{Schellekens_2005}.
An archetypal two-mode interferometer is the Mach-Zehnder configuration, where two input fields are mixed in a beam splitter, undergo  
a relative phase shift $\theta$, and are recombined in a second beam splitter. 
Two detectors, placed at the two output ports, allow the measurement of the total and relative number of particles. 
Single-atom detection with nearly unit quantum efficiency has been recently demonstrated with Bose Einstein condensates in
an optical box trap \cite{Meyrath_2005, Chuu_2005}.
From the collected data, it is possible to infer the value of $\theta$ with a certain sensitivity which mostly 
depends on the nature of the input fields \cite{Pezze_2006}.
The goal of quantum interferometry is to detect a weak external phase shift with the maximum sensitivity. 
It has been shown \cite{Giovannetti_2005} that quantum mechanics imposes a fundamental 
uncertainty on the precision $\Delta \theta$ with which the phase shift $\theta$ can be measured. 
This ultimate limit of phase sensitivity is usually discussed as the Heisenberg limit,  $\Delta \theta \sim 1/N_T$, 
being $N_T$ the total number of particles (atoms or photons) passing through the arms of the interferometer. 
Different schemes have been proposed to reach this limit \cite{Yurke_1986, Boundurant_1984, Bollinger_1996, Holland_1993}. 
In this report we focus on the Twin-Fock state first proposed in \cite{Holland_1993}:
\be \label{TwinFock}
|\psi_{inp}\rangle=|\frac{N}{2}\rangle_a |\frac{N}{2}\rangle_b.
\ee
This state provides the Heisenberg limit of phase sensitivity when it feeds the 
$a$ and $b$ input ports of a linear Mach-Zehnder interferometer. 
While it is very difficult to create the state (\ref{TwinFock}) with photons \cite{Pfister_2004}, 
Bose-Einstein condensates make possible the production of Twin-Fock states with a large number of particles through 
splitting an initial condensate using a ramping potential barrier. 
The transition from the superfluid to the Mott-insulator regime has been recently 
demonstrated in an array of wells \cite{Orzel_2001, Greiner_2002}. 
The dynamical splitting of a condensate into two parts has been experimentally studied in  \cite{Schumm_2005, Shin_2004}
and theoretically analyzed in \cite{Pezze_2005a, Meystre_2004, Pezze_2005}.   
Alternatively, the state in Eq.(\ref{TwinFock}) can be created with two condensates realized independently \cite{Saba_2005}.  

Here we analyze a crucial component of a Mach-Zehnder interferometer, namely the beam splitter which,
in quantum interferometry, transforms an uncorrelated input quantum state to an highly entangled one necessary 
to overcome the shot noise limit $1/\sqrt{N_T}$ \cite{Kim_2002, Paris_1999}. 
The beam splitter is created by a double well potential with a time-dependent barrier, 
taking into account the non-linear effects due to the 
particle-particle interaction in each condensate.
Non linearity makes the condensate dynamics highly non-trivial.
We show how non-linearity can degrade the sensitivity from the Heisenberg limit, in the 
non interacting case, toward the shot-noise limit, in the presence of strong interactions. 
We show that there is a range of values of the ratio between the non-linear interaction and the tunneling strength   
in which sub shot-noise sensitivity can still be achieved, provided the splitting is performed in the right time interval.
The non-linear interaction can be tuned, for example, with a Feshbach resonance.
A beam splitter for a Bose-Einstein condensate  
has recently been experimentally demonstrated, starting from a single condensate, using competing techniques:
in atom chips with trap deformation \cite{Schumm_2005,Shin_2005}, with matter wave Y-guide \cite{Cassettari_2000} and 
with a Bragg pulse \cite{Simsarian_2000}.
Among these, the trap deformation seems to be the most appropriate way to couple two independent condensates. 
Very stable optical double-well traps have recently been experimentally reported \cite{Albiez_2005, Shin_2004, Schumm_2005}. 
Those have found applications in the study of Josephson dynamics \cite{Albiez_2005}
and matter wave coherent splitting \cite{Shin_2004, Schumm_2005}.
The developing of a beam splitter for Bose-Einstein condensates represents a challenging
technological step toward the building of a matter wave ultrasensitive interferometer.

In section II we will introduce our beam splitter model based on a two-mode approximation of
the double-well dynamical splitting.
The relevant results of our analytical and numerical studies are presented in section IV. 
In particular we discuss how the phase sensitivity, which, in the linear limit, is given by the 
Heisenberg limit, is degraded by the non-linear particle-particle interaction 
in each condensate. 
A detailed analysis allows us to distinguish three regimes, Rabi, Josephson and Fock. 
The transition between these is characterized in sections V and VI.

\section{Beam Splitter Model}

\begin{figure}[!ht]
\begin{center}
\includegraphics[scale=0.7]{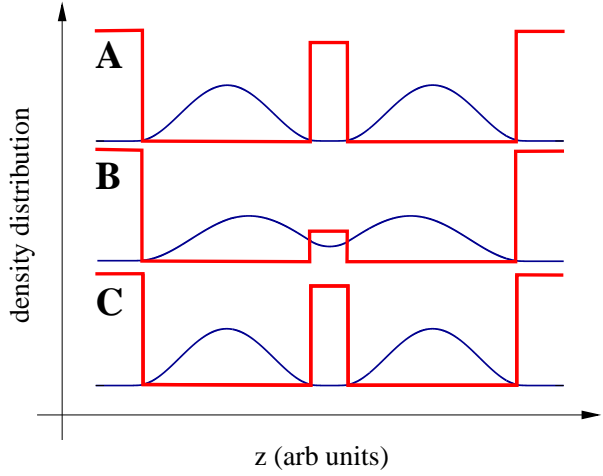}
\end{center}
\caption{\small{Schematic representation of the beam splitter for a BEC in a double square well potential. 
The red line represents the potential barrier, the blue line is the condensate density distribution 
obtained by solving a one dimensional Gross-Pitaevskii equation.
The beam splitter is realized in three steps:
A) we start with two independent condensates, B) we allow a tunneling between the two wells for a certain time $t$ 
by decreasing the height of the potential barrier separating the two condensates, and
C) we rise the potential barrier to suppress the tunneling.}}\label{wells}
\end{figure}

In the linear limit of non-interacting bosons, the 50/50 beam splitter is represented by the unitary transformation \cite{Yurke_1986}
\be \label{linearBS}
|\psi_{out} \rangle = e^{-i \frac{\pi}{2} \big(\frac{\hat{a}^{\dag}\hat{b}+\hat{b}^{\dag}\hat{a}}{2}\big)} |\psi_{inp} \rangle,
\ee 
$\hat{a}$ and $\hat{b}$ being annihilator operators for the two input ports; $|\psi_{inp} \rangle$ and $|\psi_{out} \rangle$, the 
input and output states of the beam splitter, respectively. 
A linear beam splitter for photons can be realized with a half transparent lossless mirror, while,
in the case of ions it is given by a $\pi/2$ Raman pulse \cite{Wineland_1994}. 

For interacting Bose-Einstein condensates in a double-well potential, the beam splitter transformation, 
in a two-mode model, can be written as
\be \label{nonlinearBS}
|\psi_{out} (t) \rangle = e^{-i \big(
\frac{E_c(t)}{2} \, \hat{a}^{\dag}\hat{a} \, \hat{b}^{\dag}\hat{b} 
+ K(t) \,(\hat{a}^{\dag}\hat{b}+\hat{b}^{\dag}\hat{a})
\big) t} |\psi_{inp} \rangle,
\ee
where 
\be \label{twomode}
\hat{H}= - \frac{E_c(t)}{2} \, \hat{a}^{\dag}\hat{a} \, \hat{b}^{\dag}\hat{b} 
- K(t) \, (\hat{a}^{\dag}\hat{b}+\hat{b}^{\dag}\hat{a})
\ee 
is the two-mode Hamiltonian \cite{Javanainen_1999}, 
$E_c(t)$ being the charging energy, proportional to the particle-particle interaction in each condensate, 
and $K(t)$ the coupling energy, representing the tunneling strength between the two condensates. 
The ratio between these two parameters can be controlled by tuning the interaction strength by a Feshbach 
resonance or, dynamically, by adjusting the height of potential barrier. 
In our model, a beam splitter for interacting Bose-Einstein condensates is created through
a three stage process (see Fig. (\ref{wells})),
A) we start from two independent condensates ($E_c(0) \neq 0$, $K(0)=0$),
as described by Eq.(\ref{TwinFock}); B) we allow a tunneling between the potential 
wells by decreasing the height of the potential barrier separating them  ($E_c(t) \neq 0$, $K(t) \neq 0$);
and finally C) we suppress the tunneling by raising the potential barrier. 
We recover the 50/50 linear beam splitter Eq. (\ref{linearBS}) when the interaction is switched off,  
$E_c(t)=0$, and $\int_{0}^{+\infty} \ud t \, K(t) = \pi/4$ \cite{nota2}.
In figure (\ref{wells}) we present a schematic representation of the beam splitter for BEC in a 
double square well potential.
An optical box trap with single-atom detection capability has been recently experimentally 
realized in \cite{Meyrath_2005, Chuu_2005}.
We point out that our analysis can be easily extended to condensates in a double well potential 
of arbitrary shape.

\begin{figure}[!ht]
\begin{center}
\includegraphics[scale=0.7]{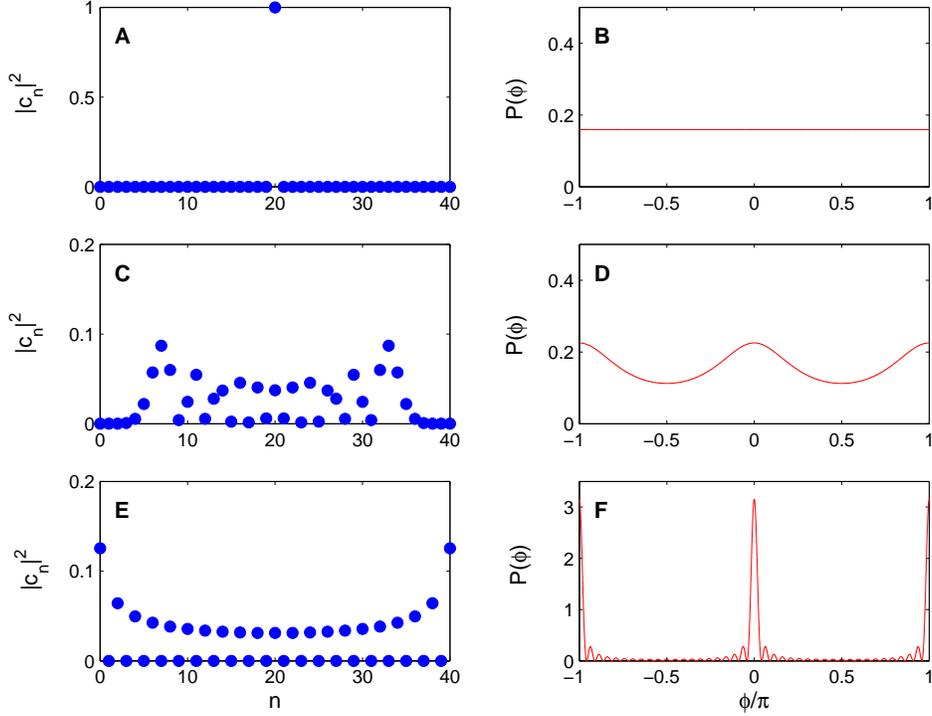}
\end{center}
\caption{\small{Relative number and phase distributions in the linear limit, plotted at different times: figures A and B, $t=0$;
figures C and D, $t=\pi/(8K)$; figures E and F, $t=\pi/(4K)$. Here $N=40$, and $K(t)$ is kept constant during the dynamics.}}\label{linear}
\end{figure}

We study the dynamics of a system of $N$ particles by projecting its quantum state onto the Fock basis
$|n\rangle \equiv |N-n\rangle_a |n \rangle_b$.
In general, the output state $|\psi_{out}\rangle$  can be written as
\be \label{psiout}
|\psi_{out} (t) \rangle=\sum_{n=0}^{N}c_n(t)|n\rangle,
\ee
where the coefficients $c_n(t)$ are given by $c_n(t)=\langle n | \psi_{out} (t) \rangle$. 
If the condensate is initially in a Twin-Fock state, $|\psi(0)\rangle=|N/2\rangle|N/2\rangle$, then we have 
$|c_{N/2}(0)|^2=1$, and $|c_n(0)|^2=0$ for $n \neq N/2$ (see Fig.(\ref{linear},A)). 
To study the phase sensitivity  we consider the operator  \cite{Sanders_1995}
\be \label{POVM}
\hat{E}(\phi)\equiv \frac{N+1}{2 \pi} |\phi \, \rangle \, \langle \phi|, 
\ee
where $|\phi \, \rangle$ are the normalized phase states $|\phi \, \rangle=\frac{1}{\sqrt{N+1}}\sum_{m=0}^{N}e^{i \, \phi \, (N/2+m)}|m\rangle$.
The operator (\ref{POVM}) has a positive spectrum and $\int_0^{2\pi}\ud \phi \,\hat{E}(\phi) = 1$. 
Therefore, it defines a positive operator value measure (POVM). For an arbitrary state $|\psi_{out}(t)\rangle$, the normalized 
probability distribution is 
\begin{eqnarray} \label{phase_distribution}
P(\phi,t)  & \equiv & \frac{N+1}{2\pi} \langle \phi| \psi_{out}(t)\rangle \langle  \psi_{out}(t)| \phi \rangle \nonumber\\
& = &\frac{1}{2\pi}\bigg|\sum_{n=0}^{N} c_n^* \, e^{i \, (N/2+n) \, \phi}\bigg|^2. \\
\nonumber
\end{eqnarray}

In the linear case, Eq.(\ref{phase_distribution}) coincides with the optimal quantum phase estimate proposed in \cite{Sanders_1995}. 
An additional linear phase shift $\theta$ simply displaces the whole
phase distribution, thus providing a phase shift-independent probability distribution.  
In order to estimate the phase sensitivity, we calculate the distance between the two first
minima on both sides of the central peak.
This method, although used by other authors \cite{Sanders_1995}, does not take into account the 
effect of the tails of the phase distribution, and it can only give qualitative results.
The phase sensitivity has to be calculated with a rigorous Bayesian analysis of quantum inference as done, in the linear case, 
and for the whole Mach-Zehnder interferometer in \cite{Pezze_2006}, where the effect of the tails of the distribution is discussed in detail. 
The initial Fock state has a flat probability phase distribution, corresponding to a complete undefined relative phase 
between the two condensates (see Fig.(\ref{linear},B)). 
The linear beam splitter, for a constant $K(t)=K$, can be studied analytically, and the results do not depend on the 
particular value of $K \neq 0$. The coefficients $c_n(t)$ are given by
\be \label{linearcn}
c_n(t)= e^{i \frac{\pi}{2} \big(\frac{N}{2}-n\big)} \sqrt{\frac{N/2! \, N/2!}{(N-n)! \, n!}} 
\Big(\sin (t K) \Big)^{n-\frac{N}{2}} \Big(\cos (t K) \Big)^{\frac{N}{2}-n}
P_{\frac{N}{2}}^{n-\frac{N}{2}, \frac{N}{2}-n}\big[\cos(2t K)\big],
\ee
where $P_{n}^{\alpha, \beta}[x]$ are the Jacobi polynomials.
In general, during the dynamics, different $c_n$ are populated, and the phase distribution develops a central peak, as shown in 
the figures (\ref{linear},C) and (\ref{linear},D), referring to the linear evolution after a time $t=\pi/(8K)$.
In figures  (\ref{linear},E) and (\ref{linear},F), we plot the $|c_n|^2$ and phase distribution after a $\pi/2$ pulse ($t=\pi/(4K)$) 
in the linear case.
As we see, the spread of the $|c_n|^2$ distribution is of the order $\sim N$ while the width of the main peak of
the phase distribution is $\sim 1/N$.

\begin{figure}[!ht]
\begin{center}
\includegraphics[scale=1]{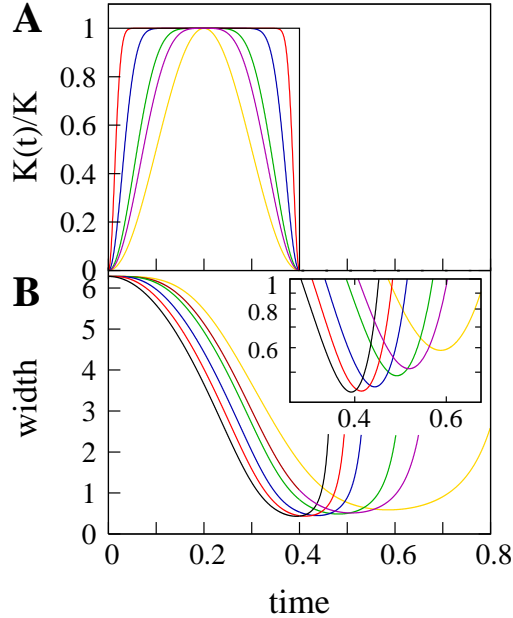}
\end{center}
\caption{\small{A) Dynamical evolution of $K(t)/K(0) = \big(1-\cos^n(t\pi/\tau)\big)$, with $\tau=0.4$ and 
for $n=2$ (yellow line), $n=4$ (violet line), $n=6$ (green line), $n=20$ (blue line), $n=100$ (red line) and $n=\infty$ (black line).
B) Width of the phase distribution as a function of time for the different $K(t)$ plotted in (A). 
The width of the phase distribution has a minimum at $t \approx 0.4$ for a sudden displacement of the double well 
(case $n=\infty$, with $K(t)/K(0) = \Theta(\tau-t))$. 
For finite values of $n$ (slow displacement) the minimum value of the width is reached 
at longer times but with larger values (lower sensitivities). 
In the insert we plot the phase distribution width as a function of time in logarithmic scale.
The best scenario is obtained for the fastest splitting. 
Here we considered $K=0.5$, $N=40$ and a constant $E_c=0.4$. Time is in units $\pi/(4 K)$.}}\label{nonlinear}
\end{figure}

In contrast with the linear case, in non-linear dynamics, the time evolutions of $K(t)$ and $E_c(t)$ play crucial roles. 
As a first approximation of the dynamical splitting, we consider a sudden displacement of the double-well at $t=0$ and 
for a certain time interval $\tau$.
In this case, we have $E_c(t)=E_c$ and $K(t)=K \Theta(\tau-t)$, where $\Theta(t)$ is the step function and $K$ is a tunable parameter.
We have checked, by 1-D Gross-Pitaevskii numerical simulations, that the parameter $E_c$ does not change significantly by displacing the 
double-wells, at least for a small overlap of the two-mode wave functions. 
This sudden displacement is the best scenario, as a slow separation of the wells will 
decrease the sensitivity, as shown in figure (\ref{nonlinear}).
In all our discussion, we develop a two-mode approximation, neglecting spurious excitations that can arise from the fast splitting of the wells.

\begin{figure}[!h]
\begin{center}
\includegraphics[scale=0.5]{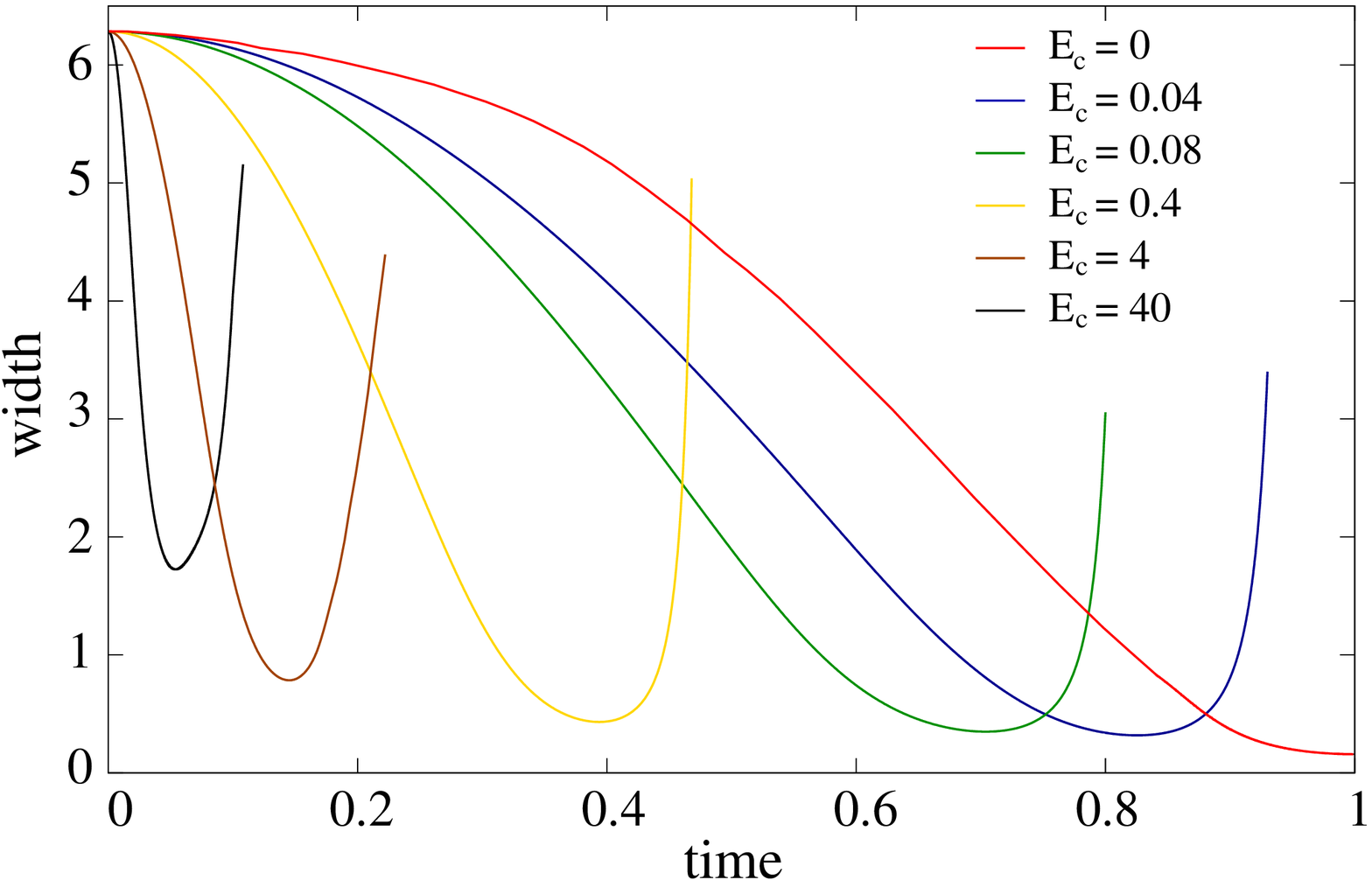}
\end{center}
\caption{\small{Width of the phase distribution Eq.(\ref{phase_distribution}) as a function of time (in units $\pi/(4 K)$) 
and for different value of the interaction energy $E_c$. The linear case ($E_c=0$) is represented by the red line. By increasing $E_c$
the minimum is attained at smaller times and at corresponding larger widths. 
We notice that, independently of $E_c$, at $t=0$ we have a flat probability distribution 
(see Fig.(\ref{linear},B)) corresponding to a phase width $2 \pi$ and complete phase uncertainty.
Here $K=0.5$ and $N=40$.}}\label{width}
\end{figure}

\section{Results}

With the formalism developed above, we now discuss the main results of our numerical and analytical study. 
First we analyze the dynamical change of the width of the phase distribution, keeping in mind that
the smaller the width, the larger the phase sensitivity.
The linear case can be studied analytically (see Eq. (\ref{linearcn})), and it is well known that the width attains its minimum 
at $t=\pi/(4 K)$, corresponding to a $\pi/2$ pulse or 50/50 beam splitter.
The linear dynamics are periodic in time with period $\pi /(2 K)$.
In figure (\ref{width}) we present the width of the phase distribution as a function of time (in units $\pi/(4 K)$) 
for different values of the charging energy $E_c$ and for fixed values of $K=0.5$ and $N=40$. 
By increasing $E_c$, the phase width reaches a minimum at smaller times and corresponding larger values. 
To these minimum of the phase distribution width there corresponds an optimal separation time to realize the non-linear beam splitter. 
In Fig. (\ref{width}) we have plotted the dynamics just after the minimum.
At longer times, the dynamics become almost chaotic, and the absolute minimum, 
together with the optimal separation time, is the one considered in the figure.

\begin{figure}[!ht]
\begin{center}
\includegraphics[scale=0.7]{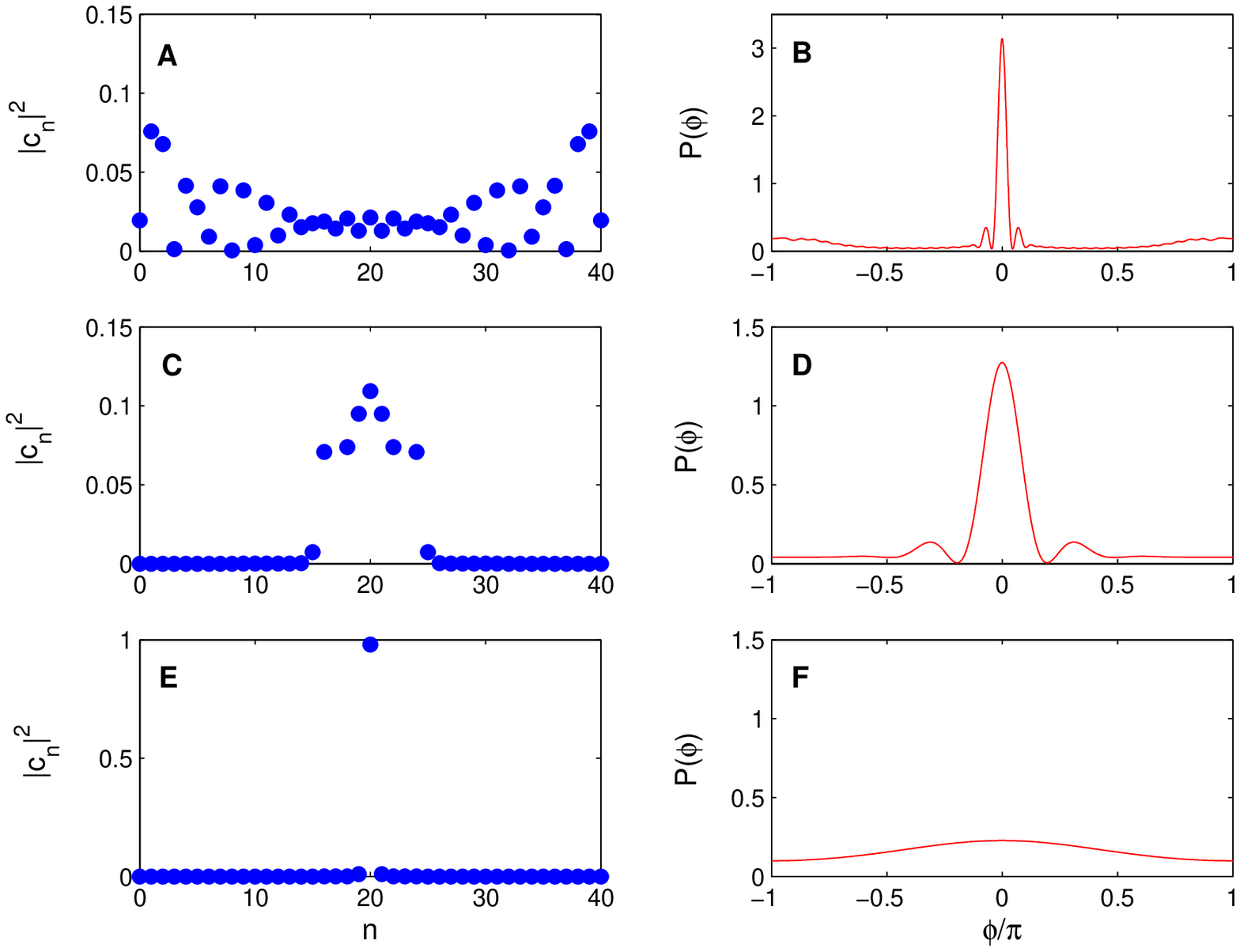}
\end{center}
\caption{\small{Relative number and phase distributions at the optimal working point (minimum phase width), for 
different values of $E_c$: A and B, $E_c=0.04$, C and D, $E_c=4$, E and F, $E_c=400$. 
These distributions can be compared with the ones in Figures (\ref{linear},E) and  (\ref{linear},F)
calculated in the linear regime, $E_c=0$, and at the optimal time $\pi/(4 K)$.
Here the parameters are the same as in Fig.(\ref{width}), $N=40$ and $K=0.5$.}}\label{minimum}
\end{figure}

In figure (\ref{minimum}) we show the relative number and phase distribution corresponding to the maximum sensitivity for given $E_c$.
It can be compared with the linear case $E_c=0$ (see Figs.(\ref{linear},E) and (\ref{linear},F)).
As shown in Fig.(\ref{minimum}), for small values of $E_c$ the phase distribution matches the linear one 
(compare Fig.(\ref{minimum},B) with  Fig.(\ref{linear},F)), and it is characterized by a narrow central peak.
We notice that the $\pi$-periodicity of the perfect linear case is lost in place of a $2\pi$-periodicity 
(this effect characterizes the presence of a non linear coupling and will be discussed in the following). 
Increasing $E_c$, the phase distribution broadens, as shown in  Fig.(\ref{minimum},D)
and eventually becomes flat, as in Fig.(\ref{minimum},F). 
This matches a loss of phase sensitivity due to the non linear particle-particle interaction in each condensate. 
The $|c_n|^2$ distribution becomes progressively narrow.
When $E_c \gg K$, for a general input state $|\psi_{inp}\rangle = \sum_{n=0}^{N} c_n |n\rangle$, we have
\be \label{grandik}
|\psi_{out}\rangle = \sum_{n=0}^{N} c_n(0)  \, e^{- i \frac{E_c}{2} n(N-n) \, t} |n\rangle.
\ee 
The dynamical evolution of each $c_n$ is simply given by an evolution of its phase, corresponding   
to a time invariance of $|c_n(t)|^2$. 
The beam splitter becomes inefficient and it does not appreciably modify the input state. 

\begin{figure}[!ht]
\begin{center}
\includegraphics[scale=0.6]{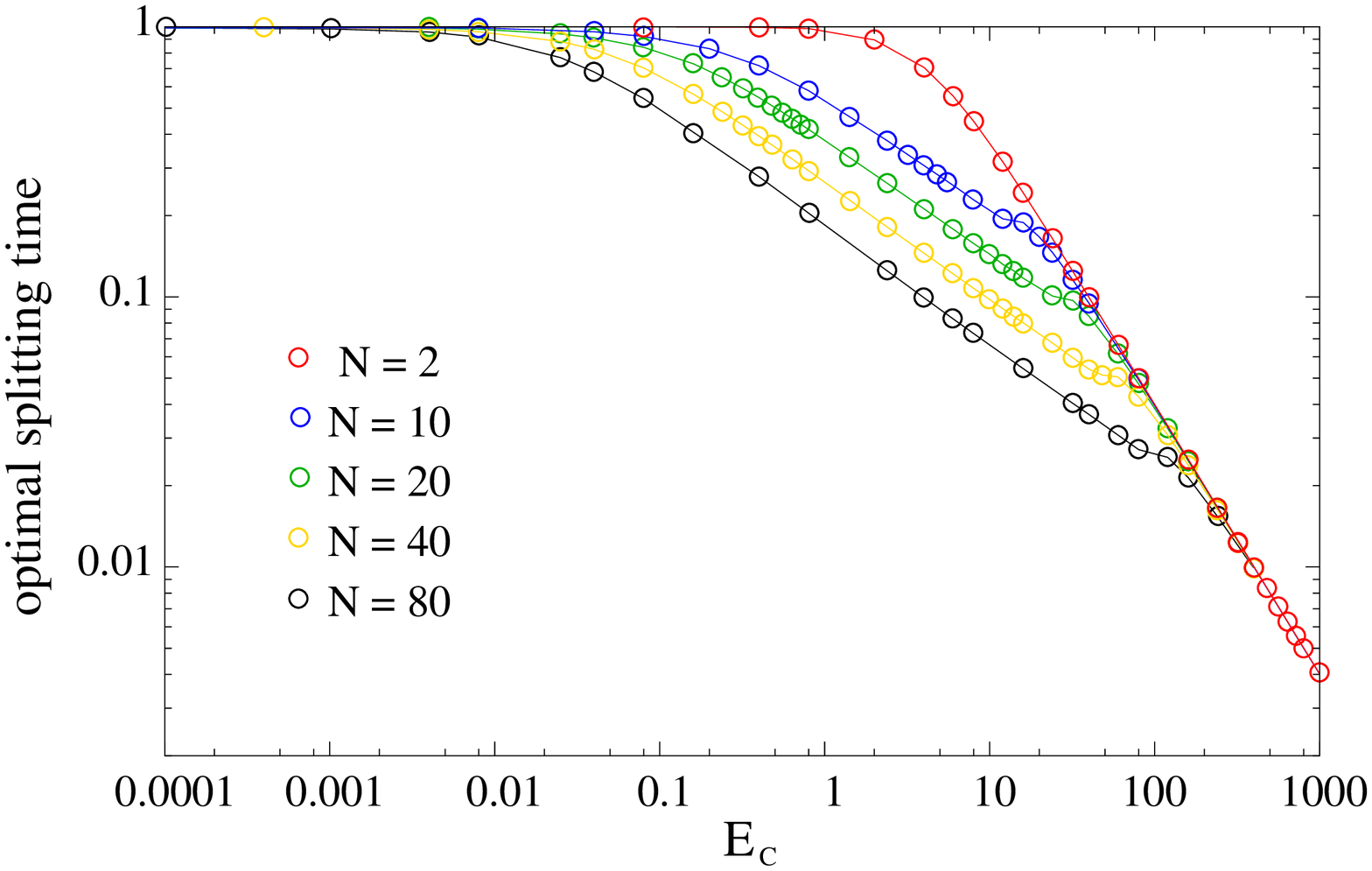}
\end{center}
\caption{\small{Optimal splitting time (in units $\pi/(4K)$) as a function of $E_c$ and for different values of $N$. 
Points are numerical results, lines are  guides to the eye. Here $K=0.5$.}}\label{timemin}
\end{figure}

In Fig. (\ref{timemin}) we present, for different values of $E_c$ and different numbers of particles,
the optimal splitting time, which defines the splitting time giving the best phase sensitivity. 
In the linear limit it is given by $t=\pi/(4 K)$ regardless of the numbers of particles.
It decreases by increasing $E_c$,
and in the limit $E_c \gg K$, when the dynamics are described by Eq.(\ref{grandik}), the optimal splitting time becomes independent of $N$. 
This effect is highlighted in the figure by the asymptotic matching of different lines corresponding to different number of particles. 

\begin{figure}[!ht]
\begin{center}
\includegraphics[scale=1]{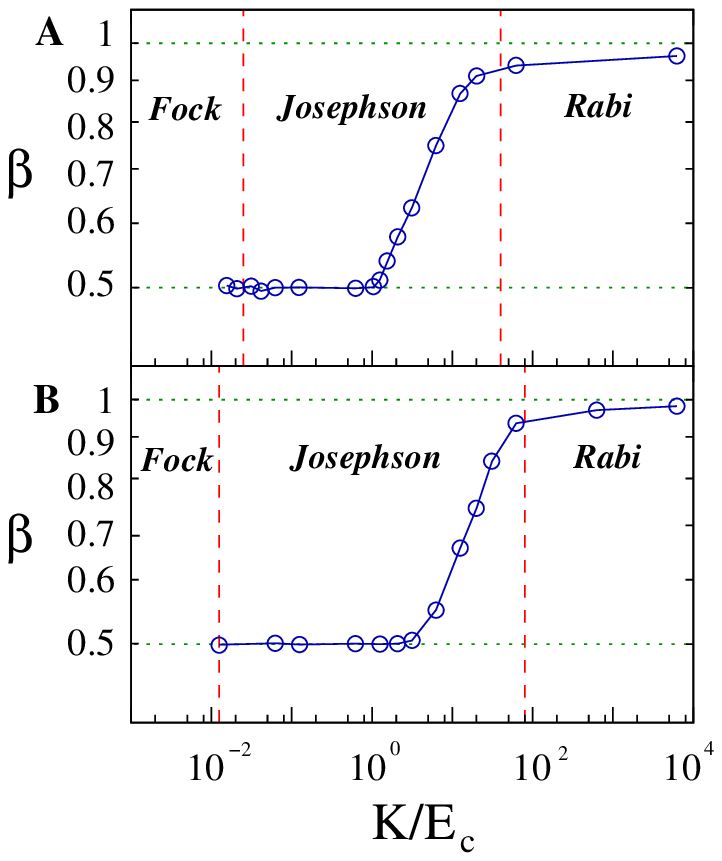}
\end{center}
\caption{\small{Scaling parameter $\beta$ (defined in Eq.(\ref{beta})) as a function of $K/E_c$. 
Points represent results of the numerical simulations, the blue line is a 
guide to the eye. The red vertical dashed lines delineate three regimes: Rabi, Josephson and Fock. 
Horizontal green dotted lines indicate the two relevant limits of quantum interferometry: the Heisenberg limit $\beta=1$, 
and the Standard Quantum Limit $\beta=0.5$.
Note that the Heisenberg Limit is reached asymptotically in the number of particles. 
Here we considered  A) $N=40$ and B) $N=80$.
The Josephson regime becomes larger when increasing the number of particles.
In the Fock regime it is not possible to define   
a scaling of the phase uncertainty since the phase distribution is almost flat 
and the phase sensitivity is $\Delta \theta \sim 2\pi$.}}\label{scaling}
\end{figure}

Asymptotically in the number of particles, we can define the phase sensitivity as
\be \label{beta}
\Delta \theta = \frac{\alpha}{N^{\beta}},
\ee
where the prefactor $\alpha$ and the scaling factor $\beta$ depend, in general, on the parameters $K$ and $E_c$. 
In figure (\ref{scaling}) we show $\beta$, as a function of $K/E_c$ and for 
different values of N ($N=40$ in Fig.(\ref{scaling},A) and $N=80$ in Fig.(\ref{scaling},B)). 
We can clearly distinguish between three regimes that characterize the two-mode dynamics.  
Following the notation introduced by Leggett \cite{Leggett_1998}, we identify the Rabi, Josephson, and Fock regimes, depending 
on the ratio $K/E_c$ (see also \cite{Pezze_2005}).   
In order to find a qualitative definition of the three regimes, 
we consider the exact quantum phase model retrieved in \cite{Anglin_2001, Pezze_2005}.
By projecting the two-mode Hamiltonian (\ref{twomode}) over the overcomplete Bargman basis \cite{Anglin_2001} we obtain the effective Hamiltonian
\be \label{EQPM}
H_{eff}(\phi, t)=- \frac{E_c}{2} \frac{\partial^2}{\partial^2 \phi} - K(t) N \cos \phi - \frac{K(t)^2}{E_c} \cos 2 \phi.
\ee
The Rabi regime is defined by $K/E_c \gg  N$. In this case the $\cos 2\phi$ term in the Hamiltonian (\ref{EQPM}) is dominant
over the  $\cos \phi$.
The effect of the $\cos 2 \phi$ term is to dynamically squeeze 
the initial flat phase distribution creating a central peak with a width at the Heisenberg limit. 
In fact, in this regime, we have $\beta \sim 1$, corresponding to sub shot-noise sensitivity. 
In particular, for $E_c=0$, the $\cos \phi$ term becomes negligible if compared with the $\cos 2 \phi$ term and the 
resulting phase distribution has period $\pi$ as noticed above (see Fig.(\ref{linear},F)). 
Physically, this is a consequence of the perfect symmetry of both the input Twin Fock state and the 50/50 
beam splitter. 
As soon as $E_c \neq 0$, the $\pi$-periodicity is lost as a consequence of the presence of the $\cos \phi$ term in Eq.(\ref{EQPM}). 
This effect can be observed in Fig.(\ref{minimum},B).\\
The Josephson regime is given by  $1/N \ll K/E_c \ll N $. 
It is the dominant regime when we increase the number of particles, keeping fixed the ratio $K/E_c$. 
As shown in figure (\ref{scaling}), the phase sensitivity decreases when $K/E_c < N$ and, in the Josephson regime, 
we recover the shot-noise limit $\beta=1/2$.
The prefactor $\alpha$ becomes progressively large.
In figures (\ref{minimum},C) and (\ref{minimum},D), we present the typical phase and  $|c_n|^2$ distributions 
in the Josephson regime (in these figures, $K/E_c=0.125$, and $N=40$): they have, to a good approximation, a Gaussian shape.\\
The Fock regime is characterized by $K/E_c \ll 1/N$ and corresponds to strong interaction. 
In the case $K/E_c \to 0$, the dynamics is described by Eq.(\ref{grandik}) 
and the phase distribution remains flat as shown in the Fig.(\ref{minimum},F). 
It is not possible to define a scaling with $N$, and we have $\beta = 0$.
In Figs. (\ref{minimum},E) and (\ref{minimum},F), we plot the narrow $|c_n|^2$ distribution and flat phase distribution
characterizing the Fock regime (in the figures  $K/E_c=0.00125$). \\
In figure (\ref{scaling}) we highlighted the three regimes discussed above (dotted vertical lines). 
As the main result of this paper we see that, even in the presence of non-linearity, there is a 
range of values of $K/E_c$, corresponding to the Rabi regime,
where it is possible to have sub shot-noise sensitivity, even in the presence of non-linear interactions.
In the following two sections, we will discuss in detail the Rabi-Josephson and Josephson-Fock transitions.

\section{Josephson-Fock transition and Self trapping}

\begin{figure}[!ht]
\begin{center}
\includegraphics[scale=1]{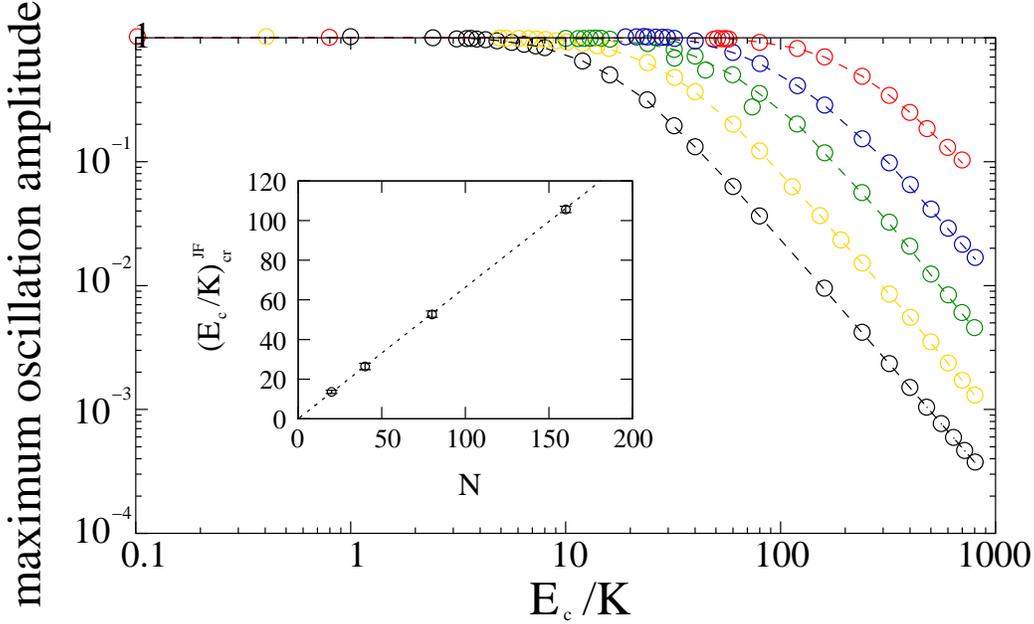}
\end{center}
\caption{\small{Maximum oscillation amplitude of $|c_{N/2}(t)|^2$, as a function of $E_c/K$ and for different numbers of particles. 
We can clearly distinguish between the selftrapped and non-selftrapped regimes. Points are numerical results, lines are guides to the eye.
Black points corresponds to $N=10$, yellow $N=20$, green $N=40$, blue $N=80$, red $N=160$.
In the insert we plot the value of $(E_c/K)_{cr}^{JF}$ corresponding to a maximum oscillation amplitude equal to 0.99, as a function of N.
We define this as the critical point between the selftrapping and non-selftrapping regimes.
Points represent numerical results with vertical bars giving the numerical uncertainty. 
The dotted line is the result of a three-mode approximation of the $c_n(t)$ dynamics, as given by Eq.(\ref{threemode}).}}
\label{selftraposc}
\end{figure}

We have observed a selftrapping effect associated with the two-mode dynamics of the system, which is different from the selftrapping 
of the Josephson oscillations \cite{Smerzi_1997,Albiez_2005}. 
It is possible to fully characterize this effect by examining the $|c_n|^2$ distribution. 
At $t=0$ the distribution is represented in Fig.(\ref{linear},A) where we have $|c_{N/2}(0)|^2=1$.  
During the dynamics, $|c_{N/2}(t)|^2$ decreased, and the other modes $|c_n(t)|^2$ with $n \neq N/2$ are populated. 
Since the energy is conserved, by monitoring the quantity $|c_{N/2}(t)|^2$, we can see how the initial energy 
dynamically distributes among all the modes. 
In particular, when  $|c_{N/2}|^2=0$, the initial energy has been completely distributed. 
In this way we can distinguish between two different behaviors: 
complete and incomplete energy distributions.
When $E_c=0$, we have that $|c_{N/2}(t)|^2$ follows a perfectly periodic motion with amplitude 1 and  period $\tau=\pi/(2K)$. 
For small values of $E_c$, we find that, initially, the $|c_{N/2}(t)|^2$ dynamics follows the linear behavior, and then 
it performs an almost chaotic motion. In this case the oscillations are still limited between 1 and $\sim 0$.
For large values of $E_c$ we observe that $|c_{N/2}(t)|^2$ oscillates between 1 and a value clearly different from 0, performing
a selftrapped dynamics. 
\begin{figure}[!ht]
\begin{center}
\includegraphics[scale=1]{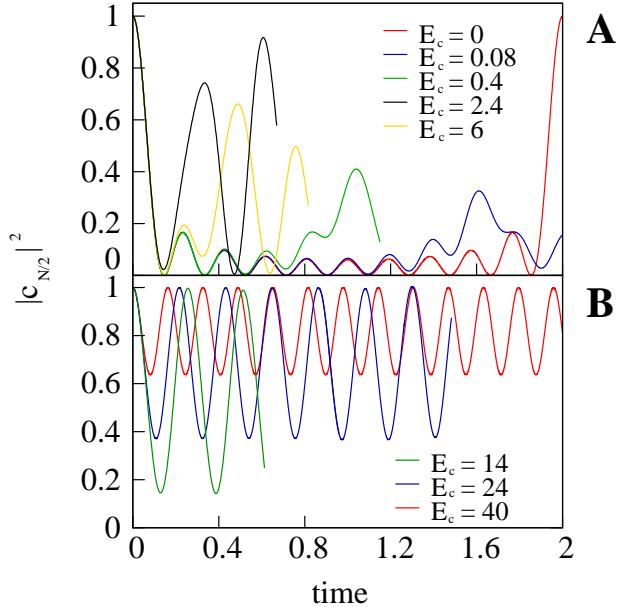}
\end{center}
\caption{\small{Time oscillation of $|c_{N/2}(t)|^2$, for different values of $E_c$ and in the selftrapping (A) and non-selftrapping (B) regimes.
Time is in units $\pi/(4K)$.}}
\label{selftrap}
\end{figure}
The two different regimes, non-selftrapped and selftrapped are shown in figure (\ref{selftrap}), 
where the time evolution of $|c_{N/2}(t)|^2$ is reported.
In Fig.(\ref{selftraposc}) we present the maximum oscillation amplitude of $|c_{N/2}(t)|^2$ for different values of $E_c/K$. 
In the non-selftrapped regime this amplitude is close to 1, while in the selftrapped region it is clearly smaller than 1. 
The selftrapped regime corresponds to an ineffective beam splitter where the two input modes are slightly mixed.
This condition matches exactly the transition between the Josephson and the Fock regimes, where the beam splitter does not 
modify the input number and phase distributions.
To find a qualitative estimate of the critical value $(E_C/K)_{cr}^{JF}$ characterizing the 
transition between the two regimes, we make a three-mode approximation of the $c_n(t)$ dynamics.
This corresponds to approximating Eq.(\ref{psiout}) as 
$|\psi_{out} (t) \rangle= \sum_{n=N/2-2}^{N/2+2}c_n(t) |n\rangle$, taking into account 
the symmetry of the distribution of $c_n$ around $n=N/2$ (See, for example Figs.(\ref{linear}) and (\ref{minimum})).
By substituting $c_{N/2}=\sqrt{p_0}e^{i\phi_0}$, and analogously for $c_{N/2+1}$ and $c_{N/2+2}$, in the limit $N \gg 1$, we obtain 
\be
\frac{E_c}{K} =2 N \,\frac{\sqrt{p_0 p_1}\cos(\phi_0-\phi_1)+\sqrt{p_1 p_2}\cos(\phi_2 - \phi_1)}{p_1 + 4 p_2}. 
\ee
With the three-mode constraint $p_0+2p_1+2p_2=1$, where we consider $p_0=0.01$, and after the direct numerical observation of $p_1$ 
for the case $N=10$, we obtain 
\be \label{threemode}
\Big(\frac{E_c}{K}\Big)_{cr}^{JF} = 0.663 \, N.
\ee
A simpler two-mode approximation would give $(E_c/K)_{cr}^{JF}=0.899\,N$.
The result is presented in Fig.(\ref{selftraposc}) which shows a perfect agreement between the numerical results (points) and the
three-mode approximation (line) given by Eq.(\ref{threemode}). 
We note that Eq.(\ref{threemode}) marks the transition between the Josephson and the Fock regimes $E_c/K \sim N$.

\section{Rabi-Josephson transition}

\begin{figure}
\begin{center}
\includegraphics[scale=1]{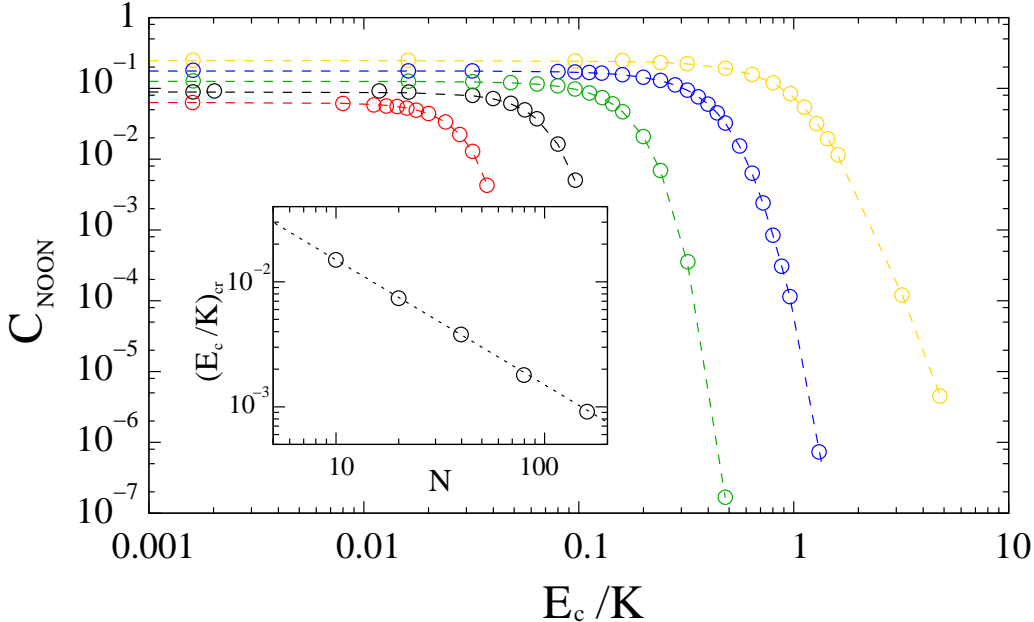}
\end{center}
\caption{\small{$C_{NOON}$, defined in equation (\ref{noonproj}), as a function of $E_c/K$. 
Points corresponds to results of numerical simulations using different number of particles: red points refer to $N=160$, 
black $N=80$, green $N=40$, blue $N=20$ and yellow $N=10$. The lines are guides to the eye.
The sharp decrease of $C_{NOON}$ defines a critical value of $E_c/K$ which is reported, as a function of the number of particles in 
the inset. The dashed line in the inset is given by Eq. (\ref{fitRJ}) and is obtained by numerical fitting.}}
\label{noonfig}
\end{figure}

To characterize the transition between the Heisenberg limit of phase sensitivity to the Standard Quantum limit, we 
introduce the entangled NOON state defined as
\be \label{noon}
|\Psi_{NOON}\rangle = \frac{|N,0\rangle+|0,N\rangle}{\sqrt{2}}.
\ee
When created after the first beam splitter of a Mach-Zehnder interferometer, this state leads to a Heisenberg 
limited phase sensitivity \cite{Bollinger_1996}. Qualitatively, this effect can be simply understood considering the 
probability distribution (\ref{phase_distribution}). 
For the NOON state (\ref{noon}) $c_0=c_N=1/\sqrt{2}$ and we have $P(\phi)=\cos^2(N\phi/2)$. 
This probability distribution is characterized by $N$ equal peaks in the interval $[-\pi, \pi]$, each
with width $2\pi/N$.  
If the projection of the state $|\psi_{out}\rangle$ (created from $|\psi_{inp}\rangle$ after the beam splitter) 
over the NOON state Eq.(\ref{noon}) is different from zero, then 
the state $|\psi_{inp}\rangle$ can be used to 
reach the Heisenberg limit of phase sensitivity in a Mach-Zehnder interferometer \cite{nota3}. 
From Eq.(\ref{psiout}), we can define the quantity 
\be \label{noonproj}
C_{NOON} \equiv  |\langle \psi_{NOON}|\psi_{out}\rangle|^2 =  |c_0(t)+c_N(t)|^2/2,
\ee
which gives the probability to obtain the NOON state, given the state $|\psi_{out}\rangle$ 
after the first beam splitter. 
In general the quantity $C_{NOON}$ depends on the parameters $K$ and $E_c$. 
For a Twin Fock input state we have $c_0(t)=c_N(t)$ and $C_{NOON}=2 |c_N(t)|^2$. 
The condition $C_{NOON} \to 0$ marks the transition from the Heisenberg limit to the Standard Quantum Limit, 
thereby characterizing the Rabi-Josephson transition.  
In figure (\ref{noonfig}) we plot the quantity  $C_{NOON}$ as a function of $E_c/K$ and for different numbers of particles. 
We notice a very fast decrease of $C_{NOON}$ after a critical point $(E_c/K)_{cr}^{RJ}$. 
In fig. (\ref{noonfig}) we plot this critical point as a function of the number of particles $N$. 
With a linear interpolation, we obtain the condition 
\be \label{fitRJ}
\Big(\frac{E_c}{K}\Big)_{cr}^{RJ} =\frac{0.15}{N},
\ee
which exactly marks the Rabi-Josephson transition ($K/E_c\sim N$).

\section{Conclusions}
Several aspects of interferometry with Bose-Einstein condensates have been discussed in the literature 
\cite{Dunningham_2002, Search_2003, Kim_1999, Pezze_2006}. 
For instance, the effect of losses and a non-linear phase shift has been described in \cite{Dunningham_2002} and \cite{Search_2003} respectively, 
and the detection efficiency, which seems to be the major obstacle toward the reach of the Heisenberg limit, has been studied in \cite{Kim_1999}.
In this paper we analyzed how the non-linear effects associated with the particle-particle interaction in each condensate 
affect the realization of a BEC beam splitter.
In particular we focused on two initially independent condensates in a Twin-Fock state. 
We showed, in a two-mode model, that the non-linear coupling decreases the interferometer phase sensitivity  from the Heisenberg limit
to the Standard Quantum Limit. 
Depending on the ratio $K/E_c$ we characterized three regions for the phase sensitivity:
the Rabi, Josephson and Fock regimes. We discussed the transitions between those regimes.
The main result of our detailed analysis of the beam splitter is that there is an interval of the parameter $K/E_c \gg N$, 
the so called Rabi regime, where sub Shot Noise sensitivity can be achieved, despite the presence of a 
non linear coupling. 
This conclusion is of interest in view of recent experiments where both the particle-particle interaction (employing a Feshbach resonance)
and the tunneling strength (tuning the potential barrier) can be appropriately controlled and changed, making 
the Rabi regime and sub-shot noise sensitivity achievable.

\section{Acknowledgement}.  
This work has been partially supported by the US Department of Energy.


\begin{thebibliography}{20}
\bibitem{nota2} We have checked numerically that in the linear limit the only condition on $K(t)$ to reach the Heisenberg limit 
is given by $\int_{0}^{+\infty} \ud t \, K(t) = \pi/4$.
We used $K(t)=K \big[1 - \big(\cos(\pi t/\tau) \big)^n \big]$ for different values of the parameters. 
If $n \to \infty$ we have $K(t) \to K \, \Theta(\tau-t)$, where $\Theta(\tau-t)$ is the usual step function, 
then we recover immediately the linear limit when $\tau=\pi/(4\,K)$.  
We also checked the constraint with a linear evolution $K(t)=K \,t \, \Theta(\tau-t) $, where $\tau=\sqrt{\pi/(2\,K)}$.
\end{thebibliography}
\end{document}